\documentclass[aps,prb,twocolumn,superscriptaddress,showpacs,showkeys,floatfix]{revtex4}

\usepackage{graphicx,color}
\usepackage{multirow,slashbox}

\graphicspath{{figs/}}
\begin{document}
\title{Elastic Bending Modulus of Single-Layer Molybdenum Disulphide (MoS$_{2}$): Finite Thickness Effect}
\author{Jin-Wu Jiang}
%    \altaffiliation{Electronic address: jinwu.jiang@uni-weimar.de}
    \affiliation{Institute of Structural Mechanics, Bauhaus-University Weimar, Marienstr. 15, D-99423 Weimar, Germany}
\author{Zenan Qi}
    \affiliation{Department of Mechanical Engineering, Boston University, Boston, Massachusetts 02215, USA}
\author{Harold S. Park}
    \altaffiliation{Electronic address: parkhs@bu.edu}
    \affiliation{Department of Mechanical Engineering, Boston University, Boston, Massachusetts 02215, USA}
\author{Timon Rabczuk}
    \altaffiliation{Electronic address: timon.rabczuk@uni-weimar.de}
    \affiliation{Institute of Structural Mechanics, Bauhaus-University Weimar, Marienstr. 15, D-99423 Weimar, Germany}
    \affiliation{School of Civil, Environmental and Architectural Engineering, Korea University, Seoul, South Korea }

%\date{22 December 2009}
\date{\today}
\begin{abstract}
We derive, from an empirical interaction potential, an analytic formula for the elastic bending modulus of single-layer MoS$_{2}$ (SLMoS$_{2}$).  By using this approach, we do not need to define or estimate a thickness value for SLMoS$_{2}$, which is important due to the substantial controversy in defining this value for two-dimensional or ultrathin nanostructures such as graphene and nanotubes.  The obtained elastic bending modulus of 9.61~{eV} in SLMoS$_{2}$ is significantly higher than the bending modulus of 1.4~{eV} in graphene, and is found to be within the range of values that are obtained using thin shell theory with experimentally obtained values for the elastic constants of SLMoS$_{2}$.  This increase in bending modulus as compared to monolayer graphene is attributed, through our analytic expression, to the finite thickness of SLMoS$_{2}$.  Specifically, while each monolayer of S atoms contributes 1.75 eV to the bending modulus, which is similar to the 1.4 eV bending modulus of monolayer graphene, the additional pairwise and angular interactions between out of plane Mo and S atoms contribute 5.84 eV to the bending modulus of SLMoS$_{2}$.
\end{abstract}

\maketitle
\pagebreak

%\section{Introduction}

Molybdenum Disulphide (MoS$_{2}$) is a semiconductor with a bulk bandgap above 1.2~{eV},\cite{KamKK} which can be further manipulated by reducing its thickness to monolayer, two-dimensional form.\cite{MakKF} This finite bandgap is a key reason for the excitement surrounding MoS$_{2}$ as compared to another two-dimensional material, graphene, as graphene is well-known to be gapless.\cite{NovoselovKS2005nat}  Because of its direct bandgap and also its well-known properties as a lubricant, MoS$_{2}$ has attracted considerable attention in recent years.\cite{WangQH2012nn,ChhowallaM} For example, Radisavljevic et al.\cite{RadisavljevicB2011nn} demonstrated the application of single-layered MoS$_{2}$ (SLMoS$_{2}$) as a transistor, which has a large mobility above 200~{cm$^{2}$V$^{-1}$s$^{-1}$}.  Several recent works have addressed the thermal transport properties of SLMoS$_{2}$ in both the ballistic and diffusive transport regimes,\cite{HuangW,JiangJW2013mos2,VarshneyV,JiangJW2013sw} while the mechanical behavior of the SLMoS$_{2}$ has also recently been investigated experimentally.\cite{BertolazziS,CooperRC2013prb1,CooperRC2013prb2,gomezAM2012}  We have also recently performed theoretical investigations considering edge effects on the Young's modulus of SLMoS$_{2}$ nanoribbons based on a recently developed Stillinger-Weber (SW) potential.\cite{JiangJW2013sw}

Besides the Young's modulus, the bending modulus is another fundamental mechanical property. For two-dimensional materials such as graphene or MoS$_{2}$, the bending modulus is important because it has been shown that the electronic properties of graphene can be strongly impacted by introducing curvature to its topology,\cite{levySCIENCE2010} which points to the important coupling between the mechanical and electrical properties in these two-dimensional materials.\cite{CastroNAH}.  The bending modulus also has strong implications for potential future flexible, or stretchable electronics applications involving SLMoS$_{2}$.

For graphene, it has been shown that the bending modulus can be analytically calculated directly from an empirical potential. Ou-Yang et al. obtained the value for the elastic bending modulus of graphene from a geometric approach.\cite{OuyangZC1997,TuZC2002} In another analytic work, the exponential Cauchy-Born rule\cite{ArroyoM2002} was applied to extract the elastic bending modulus for graphene from the Brenner empirical potential.\cite{ArroyoM2004,LuQ2009} The bending modulus value from both analytic studies shows good agreement with the experimental data.

Another important benefit of deriving the bending modulus $D$ directly from the interatomic potential, as done by~\citet{ArroyoM2004} and~\citet{LuQ2009} is that in doing so, the need to define an effective thickness $h$ of SLMoS$_{2}$, as is required from shell theory through the well-known relationship $D=E^{2D}h^2/(12(1-\nu^2))$, where $E^{2D}=Eh$ is the two-dimensional stiffness and $\nu$ is Poisson's ratio, is removed.  This is important because the precise definition of the thickness of a monolayer-thick nanostructure, dating back to nanotubes and more recently for two-dimensional graphene, has been an ongoing source of controversy\cite{yakobsonPRL1996,luPRL1997,xinPRB2000,tuPRB2002,pantanoJMPS2004,wangPRL2005,ArroyoM2004,huangPRB2006,kimJAP2011}.  In the present work, by adopting the finite crystal elasticity approach of~\citet{ArroyoM2004}, the bending modulus of SLMoS$_{2}$ is inherently thickness-independent because it is derived from a surface, and not volume energy density.

\begin{figure}[htpb]
  \begin{center}
    \scalebox{1.1}[1.1]{\includegraphics[width=8cm]{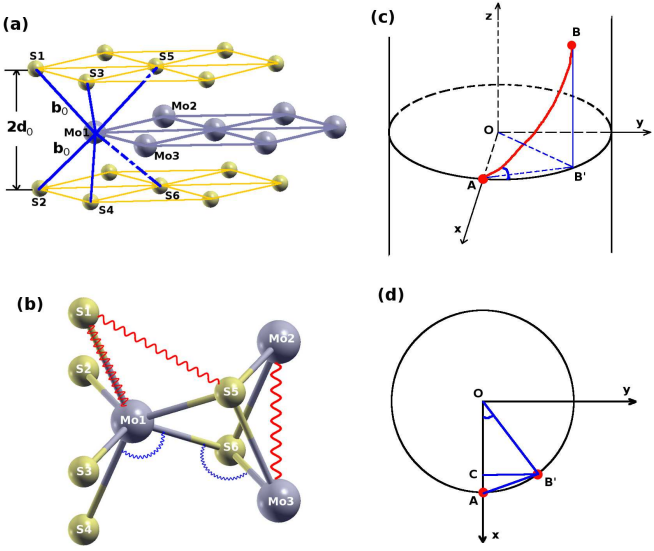}}
  \end{center}
  \caption{(Color online) The atomic configuration of the SLMoS$_{2}$. (a) and (b) are the atomic structure of Mo and S atoms. All Mo atoms are on the same plane. Atoms $S_{1}$, $S_{3}$, and $S_{5}$ are on the same atomic layer. The other three S atoms on the other atomic layer. The Mo atomic layer is sandwiched by two S atomic layers. (c) A geometrical configuration for two points A and B on a cylindrical surface. B' is the projection of B in the $xy$ plane. $\theta_{q}$ is the angle between two arcs AB and AB' on the cylindrical surface. (d) The cross-sectional view of (c).}
  \label{fig_cfg}
\end{figure}

Therefore, the objective of this work is to derive an analytic formula for the elastic bending modulus of SLMoS$_{2}$ based on our recently developed SW potential.\cite{JiangJW2013sw} The elastic bending modulus obtained for SLMoS$_{2}$ is 9.61~{eV}, which is larger than the elastic bending modulus of graphene by a factor of 7. We will demonstrate the importance of the finite thickness of SLMoS$_{2}$ in being the key factor leading to this substantial enhancement in bending modulus as compared to monolayer graphene.

\section{Geometrical constraints}

Before presenting the analytic derivation of the bending modulus, we first introduce the lattice structure for SLMoS$_{2}$, and some geometric preliminaries.  First, the crystal structure for SLMoS$_{2}$ is shown in Fig.~\ref{fig_cfg}(a), which shows that each plane of atoms takes a hexagonal structure, with two planes of S atoms and a plane of Mo atoms sandwiched in between.  This crystal structure results in three major geometric parameters as illustrated in Fig.~\ref{fig_cfg}~(a) and (b). The distance between two first-nearest-neighbor (FNN) Mo and S atoms is\cite{SanchezAM,JiangJW2013sw} $b_{0}=2.380$~{\AA}. The space between the two S atomic layers is $2d_{0}$. In our previous MD study\cite{JiangJW2013sw}, we found that three types of bond angles $\angle S_{1}Mo_{1}S_{5}$, $\angle S_{5}Mo_{1}S_{6}$, and $\angle Mo_{1}S_{5}Mo_{2}$ have the same value and the same strength.  These bond angles thus have the same bending potential energy, and the same chemical properties.  Furthermore, these angles have the same value of $2\phi_{0}$ in the undeformed SLMoS$_{2}$ configuration.

From the atomic geometry in Fig.~\ref{fig_cfg}~(a), there are the following two constraints on the variables ($b_{0}$, $d_{0}$, $\phi_{0}$):
\begin{eqnarray}\label{eq:bend1}
d_{0}=b_{0}\sin\phi_{0};\; 2d_{0}=\sqrt{3}b_{0}\cos\phi_{0}.
\end{eqnarray}
As a result, we have $\tan\phi_{0} = \frac{\sqrt{3}}{2}$ and $d_{0}  = b_{0}\sin\phi_{0}=\sqrt{\frac{3}{7}}b_{0}$, so we get the bond angle $2\phi_{0} = 81.787^{\circ}$ and the Mo-S interplane spacing $d_{0}=1.558$~{\AA}.  We denote S atoms sitting in the two planes as $S^{\pm}$. 

To investigate the bending properties of SLMoS$_{2}$, similar to graphene\cite{ArroyoM2004,LuQ2009} we homogeneously bend it into a cylindrical surface with radius $R=1/\kappa$, where $\kappa$ is the only nonzero principal curvature. Due to the homogeneous bending, the Mo atomic layer is ideally bent. However, the outer $S^{+}$ atomic layer is stretched upon bending, while the inner $S^{-}$ atomic layer is compressed. The radii of the cylinder for $S^{\pm}$ atoms are $R^{\pm}=R(1\pm\kappa d_{0})$. As a result, the tensile or compressive strain for these two S atomic layers is $\epsilon^{\pm}=\pm \kappa d_{0}$.

\section{Empirical Energy Density}

We have recently parameterized a SW potential for SLMoS$_{2}$.\cite{JiangJW2013sw} The two-body interaction takes form 
\begin{eqnarray}
V_{2}=\epsilon A\left(B\sigma^{p}r_{ij}^{-p}-\sigma^{q}r_{ij}^{-q}\right)e^{[\sigma\left(r_{ij}-a\sigma\right)^{-1}]},
\label{eq_sw2}
\end{eqnarray}
where the exponential function ensures a smooth decay of the potential to zero at the cut-off, which is key to conserving energy in MD simulations.

The three-body interaction is
\begin{eqnarray}
V_{3}=\epsilon\lambda e^{\left[\gamma\sigma\left(r_{ij}-a\sigma\right)^{-1}+\gamma\sigma\left(r_{jk}-a\sigma\right)^{-1}\right]}\left(\cos\phi_{jik}-\cos2\phi_{0}\right)^{2},
\label{eq_sw3}
\end{eqnarray}
where $2\phi_{0}$ is the angle in the undeformed configuration. There are five types of interactions in SLMoS$_{2}$ as denoted by red springs (for two-body) and blue springs (for three-body) in Fig.~\ref{fig_cfg}~(b). All SW potential parameters can be found in Ref.~\onlinecite{JiangJW2013sw}.  This potential was found to give good agreement to an experimentally-obtained phonon spectrum,\cite{JiangJW2013sw} while also yielding results for the Young's modulus of SLMoS$_{2}$ of about 229 GPa, which is within the $198.6\pm49.7$ GPa value obtained from recent experiments,\cite{BertolazziS,CooperRC2013prb1} which serves as validation of the potential's ability to accurately capture the mechanical behavior and properties, particularly within the elastic regime, or SLMoS$_{2}$.

The bending energy density within each unit cell is $W$:
\begin{eqnarray}
&&W\times S_{0}\nonumber\\
 & = & 1\sum_{\sigma=\pm}\sum_{q=FNN}^{3}V_{2}\left(r_{Mo1}^{q\sigma}\right)+\frac{1}{2}\sum_{q\in SNN}^{6}V_{2}\left(r_{Mo1-Moq}\right)\nonumber\\
&+&\frac{1}{2}\sum_{\sigma=\pm}\sum_{q\in SNN}^{6}V_{2}\left(r_{S}^{q\sigma}\right) +1\sum_{\sigma=\pm}\sum_{q=1}^{3}V_{3}\left(\theta_{Mo1}^{q\sigma\sigma}\right) \nonumber\\
 & + & 1\sum_{\sigma,\sigma'=\pm}^{'}\sum_{q=1}^{3}V_{3}\left(\theta_{Mo1}^{q\sigma\sigma'}\right)+1\sum_{\sigma=\pm}\sum_{q=1}^{3}V_{3}\left(\theta_{S^{\sigma}}^{q}\right),
\label{eq_W}
\end{eqnarray}
where $S_{0}=\frac{3\sqrt{3}}{2}c_{0}^{2}=8.423$~{\AA$^{2}$} is the area of the unit cell containing one Mo and two S atoms. For convenience, we have introduced $c_{0}  =  b_{0}\cos\phi_{0}$ as the `bond length' of the honeycomb lattice of the SLMoS$_{2}$. The honeycomb lattice is formed by Mo atoms and the projection of S atoms into the Mo atomic layer. $\sigma=\pm$ corresponds to the two S atomic layers, $r_{Mo1}^{q\sigma}$ is the bond length between atom Mo1 and its FNN S atom, which sits in the layer denoted by $\sigma$. $r_{Mo1-Moq}$ is the distance between two second-neareast-neighbor (SNN) Mo atoms, $\theta_{Mo1}^{q\sigma\sigma'}$ represents bond angles like $\angle S_{1}Mo_{1}S_{5}$ for $\sigma'=\sigma$, and $\angle S_{5}Mo_{1}S_{6}$ for $\sigma'\not=\sigma$. The prime in $\sum_{\sigma\sigma'=\pm}^{'}$ restricts $\sigma'\not=\sigma$.  An important point to emphasize in the bending energy density in Eq. (\ref{eq_W}) is that it is area, and not volume normalized, which implies that a heuristic definition of the thickness of SLMoS$_{2}$ is not required in this work.

The factor of $1/2$ in the second and the third terms is due to the fact that the two-body energy is shared between two SNN Mo or S atoms. We note that the bond $S_{5}S_{6}$ does not change during homogeneous bending, so the two-body energy association with this bond does not contribute to the bending energy density. We find that $\sum_{\sigma=\pm}=2$ in all relevant terms, because as shown above the top and bottom S atomic layers are stretched or compressed for the same amount of strain upon bending.

From the SLMoS$_{2}$ configuration, we find the following constraint due to the equilibrium of the Mo and S atoms:
\begin{eqnarray}
\frac{\partial W}{\partial r_{q}}|_{\kappa=0} = 0,
\label{eq_dWdr}
\end{eqnarray}
where $r_{q}$ is the deformed bond length.  Owing to the particular form of the SW three-body potential, we also have:
\begin{eqnarray}
\frac{\partial W}{\partial\theta}|_{\kappa=0}=0;\;\frac{\partial^{2}W}{\partial\theta\partial r_{q}}|_{\kappa=0}=0.
\label{eq_d2Wdthetadr}
\end{eqnarray}
Therefore, the SW potential predicts a zero bending modulus for graphene, because as explained in~\citet{ArroyoM2004}, the SW potential cannot describe the bending properties of planar, one-atom-thick structures.  In particular, the flexural modes, which are related to the bending modulus in graphene, will have zero energy from the SW potential.\cite{JiangJW2011bntube} However, the SW potential is able to describe the bending of SLMoS$_{2}$, which has finite thickness and non-planar covalent bonds. This point will be clearly demonstrated in the following analytic derivation of the bending modulus from the energy density $W$ in Eq. (\ref{eq_W}).

\section{Analytic Derivation of Bending modulus}

Following~\citet{ArroyoM2004}, the bending modulus can be calculated by
\begin{eqnarray}
D & = & \frac{\partial^{2}W}{\partial \kappa ^{2}}.
\label{eq_D}
\end{eqnarray}
Recalling Eqs.~(\ref{eq_dWdr}) and (\ref{eq_d2Wdthetadr}), the bending energy can also be written as 
\begin{eqnarray}
D & = & \sum_{q}\frac{\partial^{2}W}{\partial r_{q}^{2}}\left(\frac{\partial r_{q}}{\partial\kappa}\right)^{2}+\sum_{q}\frac{\partial^{2}W}{\partial\theta_{q}^{2}}\left(\frac{\partial\theta_{q}}{\partial\kappa}\right)^{2}.
\label{eq_D2}
\end{eqnarray}
This formula is substantially different from the bending modulus formula in graphene.\cite{ArroyoM2004} Specifically, the first derivative here for $r_{q}$ and $\theta_{qk}$ with respect to $\kappa$ is nonzero owing to the finite thickness of SLMoS$_{2}$. 

To calculate the bending modulus using Eq. (\ref{eq_D2}), two quantities $\frac{\partial r_{q}}{\partial\kappa}$ and $\frac{\partial\theta_{qk}}{\partial\kappa}$ are required. In the following, we will calculate these two quantities for all six terms in the energy density $W$ in Eq. (\ref{eq_W}).  Finally, while SLMoS$_{2}$ is a multi-lattice which requires an internal, or shift degree of freedom between the Mo and S planes, we have verified that, similar to monolayer graphene\cite{ArroyoM2004,LuQ2009}, the shift degree of freedom does not contribute to the bending modulus of SLMoS$_{2}$.

\textit{(1)} The first term in the energy density $W$ in Eq. (\ref{eq_W}) is of the form $V_{2}(r_{Mo1}^{q\sigma})$, which captures the pair FNN interactions between Mo and S atoms.  For the first energy term, we have $\frac{\partial^{2}W}{\partial\theta^{2}_{q}}=0$. Hence we only need to calculate $\frac{\partial r_{q}}{\partial\kappa}$. To derive this quantity, in Fig.~\ref{fig_cfg}~(c), point A represents atom Mo$_{1}$ and point B represents the projection of two S atoms (eg. S$_{5}$ and S$_{6}$) onto the Mo atomic layer. We consider the inner $S^{-}$ atom layer. From Fig.~\ref{fig_cfg}~(d), one can find the lattice vector to be:
\begin{eqnarray}
\vec{r}_{q}=\vec{AS^{-}} & = & \left(\begin{array}{c}
\left(R-d_{0}\right)\cos\kappa w_{2}\\
\left(R-d_{0}\right)\sin\kappa w_{2}\\
w_{1}
\end{array}\right)-\left(\begin{array}{c}
R\\
0\\
0
\end{array}\right)\nonumber\\
 & = & \left(\begin{array}{c}
-\frac{\kappa w_{2}^{2}}{2}Q^{2}\left(\frac{\kappa w_{2}}{2}\right)-d_{0}\cos\left(\kappa w_{2}\right)\\
w_{2}Q\left(\frac{\kappa w_{2}}{2}\right)-d_{0}\sin\left(\kappa w_{2}\right)\\
w_{1}
\end{array}\right).
\label{eq_rvec}
\end{eqnarray}
The two variables $(w_{2}, w_{1})=c_{0}(\cos\theta_{q}, \sin\theta_{q})$, where $\theta_{q}$ is the angle between the two arcs $AB$ and $AB'$ on the cylindrical surface in Fig.~\ref{fig_cfg}~(c). Eq.~(\ref{eq_rvec}) gives the lattice vector in the SLMoS$_{2}$ during bending. For $d_{0}=0$, this formula turns out to be the result of graphene, which can be obtained by the geometric approach,\cite{OuyangZC1997} or the exponential Cauchy-Born rule\cite{ArroyoM2004}. Eq.~(\ref{eq_rvec}) is actually the generalization of the geometric approach results or the exponential Cauchy-Born rule to a curved surface of finite thickness.

Using Eq.~(\ref{eq_rvec}) the first derivative of the lattice vector is
\begin{eqnarray}
\frac{\partial\vec{r}_{q}}{\partial\kappa}|_{\kappa=0} & = & \left(\begin{array}{c}
-\frac{1}{2}w_{2}^{2}\\
-d_{0}w_{2}\\
0
\end{array}\right).
\label{eq_drvec}
\end{eqnarray}
The first derivative of the bond length is
\begin{eqnarray}
\frac{\partial r_{q}}{\partial\kappa} & = & \frac{1}{r_{q}}\vec{r}_{q}\cdot\frac{\partial\vec{r}_{q}}{\partial\kappa} = -\frac{1}{2}\frac{d_{0}}{b_{0}}w_{2}^{2} = -\frac{2\sqrt{3}}{7\sqrt{7}}b_{0}^{2}\cos^{2}\theta_{q}.
\label{eq_dr}
\end{eqnarray}
This is different from the situation in monolayer graphene. We obtain a nonzero value for the first derivative of the bond length because $\vec{r}_{q}\cdot \frac{\partial\vec{r}_{q}}{\partial\kappa}\not=0$. This term is related to the inter-layer spacing $d_{0}$, which implies that this nonzero value is the result of the finite thickness of SLMoS$_{2}$. For S$^{+}$ atoms on the outer cylindrical surface, the only difference is to substitute $d_{0}$ by $-d_{0}$.

\textit{(2)} The second term in the energy density $W$ in Eq. (\ref{eq_W}) is of the form $V_{2}(r_{Mo1-Moq})$, which captures the pair SNN interactions between Mo atoms.  For the second term, we have $\frac{\partial^{2}W}{\partial\theta^{2}_{q}}=0$. It can also be shown that $\frac{\partial r_{q}}{\partial\kappa}=0$. For this derivation, in Fig.~\ref{fig_cfg}~(c), point A represents atom Mo$_{1}$ while point B represents one of its SNN atoms (eg. Mo$_{2}$). All Mo atoms are on the same cylindrical surface, so we get the lattice vector in the cylinder:
\begin{eqnarray}
\vec{r}_{q}=\vec{AB}
 & = & \left(\begin{array}{c}
-\frac{\kappa w_{2}^{2}}{2}Q^{2}\left(\frac{\kappa w_{2}}{2}\right)\\
w_{2}Q\left(\frac{\kappa w_{2}}{2}\right)\\
w_{1}
\end{array}\right),
\label{eq:term2}
\end{eqnarray}
where the two variables $(w_{2}, w_{1})=b_{Mo}(\cos\theta_{q}, \sin\theta_{q})$. $b_{Mo}=\sqrt{3}c_{0}$ is the distance between two neighboring Mo atoms in SLMoS$_{2}$. Using this formula, we find that 
\begin{eqnarray}
\frac{\partial\vec{r}_{q}}{\partial\kappa}|_{\kappa=0} & = & \left(\begin{array}{c}
-\frac{1}{2}w_{2}^{2}\\
0\\
0
\end{array}\right).
\label{eq:term2a}
\end{eqnarray}
As a result, $\vec{r}_{q}\cdot \frac{\partial\vec{r}_{q}}{\partial\kappa}=0$, leading to $\frac{\partial r_{q}}{\partial\kappa}=0$, and thus the SNN Mo atom interactions do not contribute to the bending modulus of SLMoS$_{2}$.

\textit{(3)} The third term in the energy density $W$ in Eq. (\ref{eq_W}) is of the form $V_{2}(r_{S}^{q\sigma})$, which captures the pair SNN interactions between S atoms that lie in the same plane.  For the third energy term, we have $\frac{\partial^{2}W}{\partial\theta^{2}_{q}}=0$. We consider the $S^{-}$ atom on the inner cylindrical surface. For $S^{+}$ atoms, the derivation is analogous. For this derivation, in Fig.~\ref{fig_cfg}~(c), points A and B represent two neighboring $S^{-}$ atoms (eg. S1 and S3) on the inner atomic cylindrical surface. The lattice vector is:
\begin{eqnarray}
\vec{r}_{q}=\vec{AB} & = & \left(\begin{array}{c}
-\frac{\tilde{\kappa}\tilde{w_{2}}^{2}}{2}Q^{2}\left(\frac{\tilde{\kappa}\tilde{w_{2}}}{2}\right)\\
\tilde{w_{2}}Q\left(\tilde{\kappa}\tilde{w_{2}}\right)\\
w_{1}
\end{array}\right),
\label{eq:term3a}
\end{eqnarray}
where due to the compression within the $S^{-}$ layer, we have two important relationships:
\begin{eqnarray}
\tilde{w_{2}}=\left(1-\kappa d\right)w_{2};\; \tilde{\kappa}=\frac{1}{R-d}=\kappa\left(1+\kappa d\right).
\label{eq_deffect}
\end{eqnarray}
Here, $(w_{2}, w_{1})=b_{S}(\cos\theta_{q}, \sin\theta_{q})$. $b_{S}=\sqrt{3}c_{0}$ is the distance between two neighboring S atoms in SLMoS$_{2}$.

The first derivative of the lattice vector is:
\begin{eqnarray}
\frac{\partial\vec{r}_{q}}{\partial\kappa} & = & \left(\begin{array}{c}
-\frac{1}{2}w_{2}^{2}\\
-d_{0}w_{2}\\
0
\end{array}\right).
\label{eq:term3b}
\end{eqnarray}
As a result, we get a nonzero value for the first derivative of the bond length $\frac{\partial r_{q}}{\partial\kappa}=-\frac{d_{0}}{b_{S}}w_{2}^{2}$. For $S^{+}$, an analogous derivation gives $\frac{\partial r_{q}}{\partial\kappa}=\frac{d_{0}}{b_{S}}w_{2}^{2}$.

\textit{(4)} The fourth term in the energy density $W$ in Eq. (\ref{eq_W}) is of the form $V_{3}(\theta_{Mo1}^{q\sigma\sigma})$, which captures the three-body (angular) interactions between Mo and two S atoms in the same plane.  For the fourth energy term, $\frac{\partial^{2} W}{\partial r_{q}^{2}}=0$. We consider $S^{-}$ atoms. From the definition $\cos\phi = \hat{n}_{i}\cdot\hat{n}_{j}$, where $\hat{n}_{i}$ and $\hat{n}_{j}$ are two unit vectors for the two bonds forming the angle $\phi$, we get 
\begin{eqnarray}
\frac{\partial}{\partial\kappa}\cos\phi & = & \frac{3}{14}\frac{d_{0}c_{0}^{2}}{b_{0}^{2}}\left(5-6\cos2\theta_{q}\right).
\label{eq:term4}
\end{eqnarray}

\textit{(5)} The fifth term in the energy density $W$ in Eq. (\ref{eq_W}) is of the form $V_{3}(\theta_{Mo1}^{q\sigma\sigma'})$, which captures the three-body (angular) interactions between Mo and S atoms in different layers.  For the fifth energy term, $\frac{\partial^{2} W}{\partial r_{q}^{2}}=0$. We also have $\frac{\partial}{\partial\kappa}\cos\phi=0$, because homogeneously bending SLMoS$_{2}$ results in the distance between two S atomic layers being unchanged, i.e. the bond length $S_{5}S_{6}$ is unchanged.  Thus, the angular three-body interactions between an Mo atom and two S atoms in different planes does not contribute to the bending modulus of SLMoS$_{2}$.

\textit{(6)}  The sixth term in the energy density $W$ in Eq. (\ref{eq_W}) is of the form $V_{3}(\theta_{S^{\sigma}}^{q})$, which captures the three-body (angular) interactions between S atoms in the same plane. For the sixth energy term, $\frac{\partial^{2} W}{\partial r_{q}^{2}}=0$. We consider the $S^{-}$ atom on the inner cylindrical surface. For this derivation, in Fig.~\ref{fig_cfg}~(c), point A represents the projection of two S atoms (eg. $S_{5}$ and $S_{6}$), and point B represents the Mo atom (eg. Mo1). The lattice vector and its derivatives are:
\begin{eqnarray}
\vec{r}_{q}=\vec{S^{-}B}=\left(\begin{array}{c}
\frac{\kappa w_{2}^{2}}{2}Q^{2}\left(\frac{\kappa w_{2}}{2}\right)+d_{0}\\
w_{2}Q\left(\kappa w_{2}\right)\\
w_{1}
\end{array}\right)\\ \nonumber
\frac{\partial\vec{r}_{q}}{\partial\kappa}=\left(\begin{array}{c}
-\frac{1}{2}w_{2}^{2}\\
0\\
0
\end{array}\right);\; \frac{\partial r_{q}}{\partial\kappa}=-\frac{1}{2}\frac{d_{0}}{b_{0}}w_{2}^{2},
\label{eq:term6a}
\end{eqnarray}
where $(w_{2}, w_{1})=c_{0}(\cos\theta_{q}, \sin\theta_{q})$. The derivative of the angle is:
\begin{eqnarray}
\frac{\partial}{\partial\kappa}\cos\phi & = & -\frac{3}{7}\times\frac{d_{0}}{b_{0}^{2}}c_{0}^{2}\left(1-\frac{1}{2}\cos2\theta_{q}\right).
\end{eqnarray}
We have established above that there are two terms (terms 2 and 5) in the energy density $W$ in Eq. (\ref{eq_W}) that do not contribute to the bending modulus of SLMoS$_{2}$.  We now evaluate the relative contributions of the other four terms using Eq. (\ref{eq_D2}) to obtain the bending modulus of SLMoS$_{2}$:

\begin{eqnarray}
D_{term1} & = & \frac{2}{S_{0}}\left[\frac{27}{49\times14}b_{0}^{4}V_{2}''\left(b_{0}\right)\right]=3.09 \\
D_{term3} & = & \frac{2}{S_{0}}\left[\frac{9}{8}d_{0}^{2}b_{S}^{2}V_{2}''\left(b_{S-S}\right)\right]=3.49 \nonumber\\
D_{term4} & = & \frac{2}{S_{0}}\left[\frac{12\times27\times43}{7\times49\times49}b_{0}^{2}V_{3}''\left(\cos\theta\right)\right]=2.75 \nonumber\\
D_{term6} & = & \frac{2}{S_{0}}\left[\frac{27\times54}{7\times49\times49}b_{0}^{2}V_{3}''\left(\cos\theta\right)\right]=0.29 \nonumber\\
D&=&D_{term1}+D_{term3}+D_{term4}+D_{term6}=9.61 {\rm eV}\nonumber\\ 
\label{eq_D3}
\end{eqnarray}
where the second derivatives are $V_{2}\left(r_{0}\right)'' = \frac{\partial^{2}V_{2}\left(r\right)}{\partial r^{2}}|_{r=r_{0}}$ and $V_{3}'' = \frac{\partial^{2}V_{3}\left(\cos\phi\right)}{\partial\left(\cos\phi\right)^{2}}|_{\phi=2\phi_{0}}$.
%\begin{eqnarray}
%D & = & \frac{1}{S_{0}}\{2\times\left[\frac{27}{49\times14}b_{0}^{4}V_{2}''\left(b_{0}\right)\right]+2\times\left[\frac{9}{8}d^{2}b_{S}^{2}V_{2}''\left(b_{S-S}\right)\right]\nonumber\\
% & + & 2\times\left[\frac{12\times27\times43}{7\times49\times49}b_{0}^{2}V_{3}''\left(\cos\theta\right)\right]\nonumber\\
%&+&2\times\left[\frac{27\times54}{7\times49\times49}b_{0}^{2}V_{3}''\left(\cos\theta\right)\right]\}\\\nonumber
% & = & 3.09+3.49+2.75+0.29\\\nonumber
% & = & 9.61 {\rm eV},
%\label{eq_D3}
%\end{eqnarray}

For the cycle summation $\sum_{q}$, we have used some trigonometric summation identities. If $\theta_{q} = \theta_{1} + (q-1)\frac{2\pi}{3}$, we have $\sum_{q=1}^{3}\cos\theta_{q} = 0$, $\sum_{q=1}^{3}\cos^{2}\theta_{q} = \frac{3}{2}$, and $\sum_{q=1}^{3}\cos^{4}\theta_{q} = \frac{9}{8}$. If $\theta_{q} = \theta_{1} + (q-1)\frac{\pi}{3}$, we have following trigonometric identities: $\sum_{q=1}^{6}\cos\theta_{q} = 0$, $\sum_{q=1}^{6}\cos^{2}\theta_{q} = 3$, and $\sum_{q=1}^{6}\cos^{4}\theta_{q} = \frac{9}{4}$.

From the final results in Eq.~(\ref{eq_D3}), we find the bending modulus of SLMoS$_{2}$ to be 9.61 eV, which is about 7 times larger than the value of 1.4 eV for monolayer graphene\cite{LuQ2009}.  Furthermore, we can clearly demonstrate that this difference arises due to the finite thickness effect, or the fact that SLMoS$_{2}$ actually contains three planes of atoms.  Specifically, Eq. (\ref{eq_D3}) shows that nearly 36\% of the bending modulus, or 3.49 eV, arises from the contribution of the two-body SNN interactions between S atoms that lie on the same plane.  Because there are two planes of S atoms in SLMoS$_{2}$, we find that each plane of S atoms contributes about 1.75 eV to the bending modulus.  This value is similar to the 1.4 eV value for monolayer graphene\cite{LuQ2009}.

However, due to the three planes of atoms, SLMoS$_{2}$ receives additional, out of plane contributions to its bending modulus.  Specifically, the FNN Mo-S interactions contribute about 3.09 eV, or 32\% of the total bending modulus, while the three-body (angular) Mo-S interactions between Mo and S atoms on the same plane, contributes 2.75 eV, or about 29\% of the bending rigidity.  Not surprisingly, the three-body interactions between S atoms on different planes contributes only 0.29 eV, or about 3\% to the total bending modulus.  Overall, the two-body (pair) terms contribute about 6.58, or 68\% of the total bending rigidity.  This means that the angular (three-body) contribution in SLMoS$_{2}$ of about 29\% is smaller than the 41\% contribution that the dihedral angles were found to make to the bending modulus in monolayer graphene\cite{LuQ2009}.

%the two-body interactions have important contribution to the bending modulus. This is due to the finite thickness in the SLMoS$_{2}$, which leads to tension/compress within the outer/inner S atomic layers. In other words, the bending modulus contributed by the two-body interaction actually originates in the bending-induced strain effect, which contributes about 68\% for the total bending modulus. This is also the reason why the bending modulus here is much larger than the bending modulus in graphene (~1.2~{eV}).\cite{OuyangZC1997,ArroyoM2004}

\begin{figure}[htpb]
  \begin{center}
    \scalebox{1.0}[1.0]{\includegraphics[width=8cm]{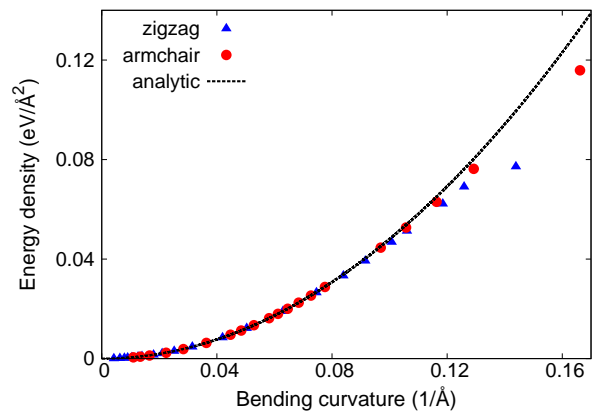}}
  \end{center}
  \caption{(Color online) Strain energy density versus the bending curvature in SLMoS$_{2}$ from molecular mechanics method for zigzag (triangles, blue online) and armchair (circles, red online) directions. The analytic result, $W=D\kappa^{2}/2$, is shown by the dashed line. Deviations between molecular mechanics and analytic results are due to nonlinearity at large bending curvature.}
  \label{fig_energy_density}
\end{figure}

To validate the analytic results, we compute the bending modulus of SLMoS$_{2}$ using the same SW potential using the molecular mechanics method. Fig.~\ref{fig_energy_density} shows the strain energy density for SLMoS$_{2}$ nanotubes. The tubes are obtained by rolling up the SLMoS$_{2}$ into a cylindrical structure, with the middle Mo atomic layer purely bent. Both armchair and zigzag tubes are calculated.  The energy is calculated for this ideally rolled up tube structure without optimization (energy minimization), because the optimization is not considered in the above analytic derivation. As we have pointed out above, the optimization of the shift degree of freedom between the Mo and S planes of atoms does not contribute to the bending modulus. However, the optimization of the whole unit cell (with one Mo and two S atoms) can slightly decrease the total energy of the system, and represents a more accurate value.  Our analytic value of 9.61 eV is about 16\% larger than the value (8.03~{eV}) obtained from the MM method with optimization and relaxation of all degrees of freedom. The dashed line in Fig.~\ref{fig_energy_density} denotes the analytic result, $W=D\kappa ^{2} /2$, with $D=9.61$~{eV}.  Good agreement is observed between the analytic result and the numerical data for curvature smaller than 0.12. Some obvious discrepancy appears for curvature larger than 0.12, which is due to the neglect of nonlinear terms in the analytic derivation.  It should be noted that for graphene, the analytic result agrees with the molecular mechanics calculation up to a curvature value around 0.25,\cite{LuQ2009} which is much larger than the value of 0.12 reported here. This is quite reasonable, considering the finite thickness and more complicated tri-layer configuration in SLMoS$_{2}$. 

Finally, we compare our result with those that can be obtained by taking recent experimental measurements for the elastic properties of SLMoS$_{2}$, and using them in the classical bending modulus expression for thin elastic structures, $D=E^{2D}h^2/(12(1-\nu^2))$.  To do so, we note that recently,~\citet{BertolazziS} have found $E^{2D}=180\pm60$ N/m for SLMoS$_{2}$, while~\citet{CooperRC2013prb1} found $E^{2D}=130$ N/m for SLMoS$_{2}$.  Given those values, we consider $E^{2D}$ to range from 120-240 N/m, while taking $v=0.29$\cite{CooperRC2013prb1}, and the thickness $h=2\times1.558=3.116$~\AA.  Taking these values gives an experimental range for the bending modulus $D$ from 6.62 to 13.24 eV.  Our obtained value of 9.61 eV clearly fits into this range.

\section{conclusion}
In conclusion, we derived an analytic formula for the elastic bending modulus of the SLMoS$_{2}$, which does not require the definition of a thickness for SLMoS$_{2}$. The obtained elastic bending modulus is 9.61~{eV} for SLMoS$_{2}$, which is significantly larger than the elastic bending modulus of graphene, is found to be within the range of values that are obtained using thin shell theory with experimentally obtained values for the elastic constants of SLMoS$_{2}$. It is found that the finite thickness of the SLMoS$_{2}$ plays a key role in determining its bending properties.  Specifically, while each monolayer of S atoms has a bending rigidity (1.75 eV) similar to that of monolayer graphene (1.4 eV), the additional pairwise and angular interactions between Mo and S atoms contributes 5.84 eV to the bending modulus of SLMoS$_{2}$.

\textbf{Acknowledgements} The work is supported by the German Research Foundation (DFG).  HSP and ZQ acknowledge the support of NSF-CMMI 1036460.  All authors acknowledge helpful discussions with Prof. Marino Arroyo.

%\bibliographystyle{aipnum4-1}
%\bibliography{biball}

\end{document}